\documentclass[preprint,12pt]{elsarticle}
\usepackage{amssymb}
\usepackage{amsmath}
\journal{Computer Methods and Programs in Biomedicine }
\usepackage[english]{babel}
\usepackage[nottoc]{tocbibind}
\usepackage{parskip}
\usepackage{setspace}
\usepackage{graphicx}
\usepackage{booktabs}
\usepackage{array}  
\usepackage{makecell}
\usepackage{multirow}
\usepackage{float}
\usepackage{ragged2e}
\usepackage{caption}
\begin{document}

\begin{frontmatter}

\author[aff1]{Hafza Eman}
\ead{23-MS-CSc-1@students.uettaxila.edu.pk}

\author[aff1]{Furqan Shaukat}
\ead{furqan.shoukat@uettaxila.edu.pk}

\author[aff2]{Muhammad Hamza Zafar}
\ead{muhammad.h.zafar@uia.no}

\author[aff3,aff4]{Syed Muhammad Anwar}
\ead{sanwar@childrensnational.org}

\address[aff1]{Faculty of Electrical and Electronics Engineering, 
               University of Engineering and Technology, Taxila, Pakistan}

\address[aff2]{Department of Engineering Sciences, 
               University of Agder, Grimstad, Norway}

\address[aff3]{Sheikh Zayed Institute for Pediatric Surgical Innovation, 
               Children’s National Hospital, Washington DC, USA}

\address[aff4]{School of Medicine and Health Sciences, 
               George Washington University, Washington DC, USA}

\title{EMeRALDS: Electronic Medical Record Driven Automated Lung Nodule Detection and Classification in
Thoracic CT Images \thanks{Supported by Higher Education Commission Pakistan under NRPU Project No:17019}}

\begin{abstract}
\textbf{Objective:}  
Lung cancer is a leading cause of cancer-related mortality worldwide, primarily due to delayed diagnosis and poor early detection. This study aims to develop a computer-aided diagnosis (CAD) system that leverages large vision–language models (VLMs) for the accurate detection and classification of pulmonary nodules in computed tomography (CT) scans.\\
\textbf{Methods:}
We propose an end-to-end CAD pipeline consisting of two modules: (i) a detection module (CADe) based on the Segment Anything Model 2 (SAM2), in which the standard visual prompt is replaced with a text prompt encoded by CLIP (Contrastive Language–Image Pretraining), and (ii) a diagnosis module (CADx) that calculates similarity scores between segmented nodules and radiomic features. To add clinical context, synthetic electronic medical records (EMRs) were generated using radiomic assessments by expert radiologists and combined with similarity scores for final classification. The method was tested on the publicly available LIDC-IDRI dataset (1,018 CT scans).\\
\textbf{Results:}
The proposed approach demonstrated strong performance in zero-shot lung nodule analysis. The CADe module achieved a Dice score of 0.92 and an IoU of 0.85 for nodule segmentation. The CADx module attained a specificity of 0.97 for malignancy classification, surpassing existing fully supervised methods.\\
\textbf{Conclusions:}
The integration of VLMs with radiomics and synthetic EMRs allows for accurate and clinically relevant CAD of pulmonary nodules in CT scans. The proposed system shows strong potential to enhance early lung cancer detection, increase diagnostic confidence, and improve patient management in routine clinical workflows. 
\end{abstract}

\begin{graphicalabstract}
\includegraphics[width=1\textwidth]{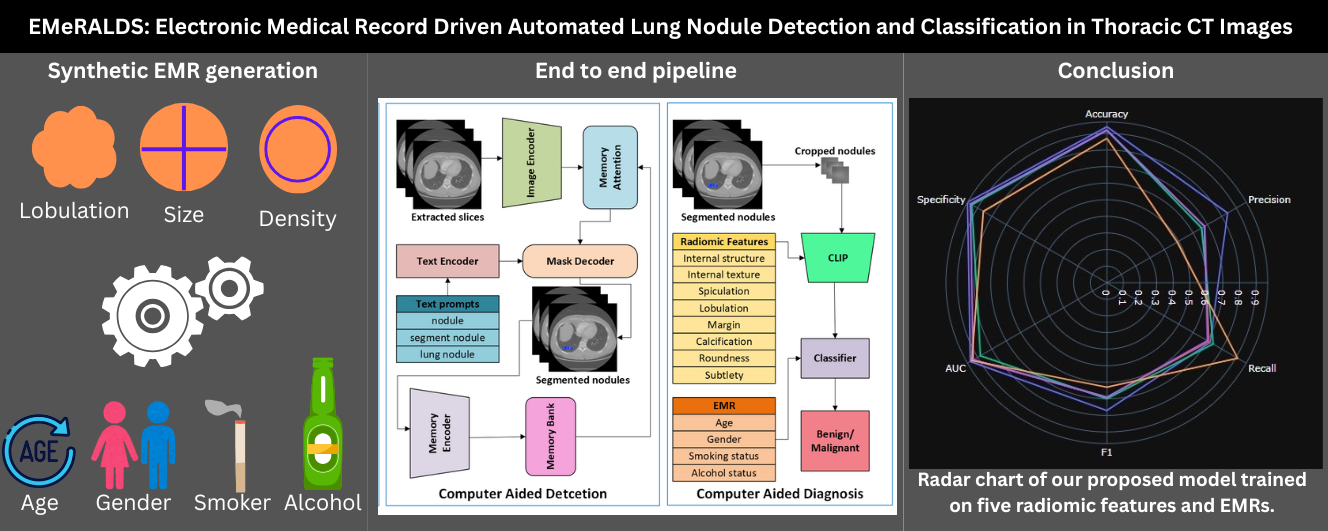}
\end{graphicalabstract}

\begin{highlights}

\item Integrating radiomic features and EMR significantly improves diagnostic performance. 
\item Unified pipeline detects and classifies lung nodules in CT scans. 
\item SAM2 with text prompts achieves zero-shot nodule segmentation.
\item Synthetic EMRs boost the accuracy of malignancy classification.
\item The model outperforms benchmarks on standard performance metrics.

\end{highlights}


\end{frontmatter}

\section{Introduction}
Lung cancer remains one of the most common cancers worldwide, accounting for approximately 20 million new cases and 9.7 million deaths in 2022~\cite{bray2024global}. 2,001,140 new cancer cases and 611,720 deaths were reported in the United States in 2024 ~\cite{siegel2024cancer}. In the year 2024, lung cancer was responsible for about 18\% of all cancer-related deaths, resulting in an estimated 1.8 million deaths ~\cite{siegel2024cancer}. The situation in developing countries is even worse, with a 63.1\% incidence rate and 62.9\% mortality rate in Asia~\cite{GLOBOCAN}. The 5-year survival rate for lung cancer is only 25\%~\cite{siegel2024cancer}. These statistics highlight severity of lung cancer and the importance of an automated system for early detection and diagnosis. Early detection of lung cancer can improve patient outcomes and reduce mortality rates~\cite{borg2024introduction}. Despite advances in medical imaging and cancer treatments, delayed diagnosis remains a challenge, leading to poor survival rates.

Medical imaging techniques such as computed tomography (CT), magnetic resonance imaging (MRI), and X-rays are important for the detection and diagnosis of cancer~\cite{LungCancerScreening,LungCancerResearch}. Among these, CT scans are mostly used because of their high resolution, which allows better visualization of lung tissues~\cite{LungCancerScreening}. Lung nodules are the primary identification of lung cancer. These nodules can be detected through computer-aided systems (CAD), which help clinicians in analyzing medical images to identify nodules and categorize them as benign or malignant ~\cite{gao2024artificial}. Computer-aided systems typically work in two stages: Computer-Aided Detection (CADe) to detect and segment the nodules from the CT scan, and Computer-Aided Diagnosis (CADx) to classify detected nodules as benign and malignant.

Computer-aided detection involves a medical image segmentation task to detect and segment lung nodules~\cite{litjens2017survey} accurately. Traditional image processing techniques such as thresholding, region-based methods, and edge detection have limitations when handling the complexity of medical images~\cite{litjens2017survey}. Recent advances in artificial intelligence (AI) have facilitated the detection and classification of lung nodules in medical imaging tasks~\cite{de2023artificial}. Deep learning and machine learning play a valuable role in developing systems to support this process~\cite{niu2022unsupervised,gedam2024hybrid,urrehman2024effective}; however, these models often depend on labeled data and face challenges with generalization. However, when processing high-resolution volumetric data, these models can be computationally expensive and resource-intensive.

Recently, the development of foundation models has enabled a dynamic shift in application of AI to healthcare. These are large-scale and generalizable models that are trained on diverse, large-scale datasets, which enables them to be applied to various downstream applications. These models have demonstrated notable performance in medical imaging tasks, especially in lung nodule detection and classification. The development of transformer-based models, such as Vision Transformer (ViT)~\cite{dosovitskiy2021imageworth16x16words}, captures long-range dependencies that improve segmentation results. Large Vision Language Models (VLM)~\cite{zhang2024vision} are trained in vast and diverse datasets, enabling them to generalize well on unseen tasks, especially those that are helpful in the healthcare community for downstream medical tasks~\cite{qin2022medical}. 

Among these, Segment Anything Model (SAM)~\cite{kirillov2023segment} has performed well in different real-world segmentation tasks. SAM is a promptable foundation model capable of zero-shot segmentation across diverse domains. Even without task-specific retraining, SAM's strong generalization capability makes it highly suitable for medical imaging, where labeled datasets are often limited. MedSAM~\cite{ma2024segment} is an adaptation of SAM specifically fine-tuned for medical images. It improves segmentation accuracy for medical images, including anatomical structures as well as lesions. SAM-Med2D~\cite{cheng2023sammed2d} further enhances SAM’s performance on 2D medical scans, demonstrating strong generalization across different modalities (CT, MRI, X-ray). Another study ~\cite{fully-automated} used MedSAM in a fully automated manner by replacing the model's visual prompts with text prompts, but the model was used on 2D slices, which causes the loss of spatial information of a CT scan. 

The improved version of the original SAM model, Segment Anything Model 2 (SAM2)~\cite{ravi2024sam}, is six times faster and more accurate. SAM2 can automate the lung nodule segmentation task using visual and text prompts, minimizing the need for a large-scale annotated dataset. Another VLM, Contrastive learning image pretraining (CLIP)~\cite{clip} enables image-text understanding to allow a model to segment images based on text prompts such as 'lung nodule'. CLIP has shown exceptional performance in matching images and texts by learning visual semantic relationships. CLIP allows the model to learn patterns and compute similarity scores based on semantic understanding rather than just relying on pixel-level patterns. Contrastive learning offers better generalization than traditional deep learning models. 

A recent study~\cite{noduleclip} combines U-Net-based 3D nodule segmentation with nodule-clip for the classification phase. This hybrid approach combines 3D U-Net with contrastive features to improve nodule localization and malignancy classification. A similar study~\cite{chexpert} adapted CLIP on CheXpert through fine-tuning that better utilizes multi-label image-report pairs. Although the model showed promising results, it still faces challenges while handling the multi-labeled nature of data. 

The second phase of medical image analysis in CAD systems is diagnosis, which focuses on determining the nature of detected nodules. CADx plays an important role by performing a classification task, which involves classifying detected nodules into benign and malignant. Machine learning and deep learning models, particularly CNN, have been extensively trained on large datasets of medical images for the diagnosis of lung cancer. However, despite significant advancements, achieving high accuracy remains a challenge due to class imbalance, limited medical data availability, and data variability. 

The advancements in VLMs and foundation models have shown significant potential in addressing these limitations. Unlike traditional deep learning techniques, which heavily rely on large-scale labeled medical datasets, VLMs utilize multimodal pretraining to reduce reliance on extensive annotations. For instance, CLIP can align visual and textual representations, which makes it more suitable to learn generalized representations that can be adopted to medical imaging tasks with limited labeled data. Additionally, foundational models can handle class imbalance by transferring knowledge from diverse domains and allow them to be used in a zero-shot or a few-shot manner. 

Recent studies highlight the potential of large vision language models and foundation models in medical images. For instance, a study~\cite{fully-automated} utilized the potential of CLIP along with radiomic features, aligning textual and visual representations for classifying nodules as benign and malignant. The proposed model achieved a sensitivity of 0.86. However, the model relies on visual nodule patches and radiomic features only. The clinical context of the nodules is not present, which is a critical requirement for clinical deployment. Nodule-CLIP~\cite{noduleclip} is another contrastive learning framework that aligns 3D CT image features with nodules' attributes to improve the classification. Although the model showed promising results, the clinical context is still missing, affecting real-world performance. 

Medical images alone lack the clinical context that can be helpful in precise diagnosis. While CT scans provide detailed anatomical information, additional Electronic Medical Records (EMR) such as age, gender, smoking history, and alcohol history can significantly improve diagnosis accuracy~\cite{BENASSULI201531}. EMR provide contextual information that can enhance the nodule classification task. Foundation models eliminate the need for extensive training, making them more efficient for clinical deployment. However, key challenges remain, which include generalization across datasets, computational efficiency, EMR heterogeneity, and interpretability. 
This study introduces a CAD system that consists of two components. \textbf{1)Computer Aided Detection(CADe)} that performs detection and segmentation of nodules from CT scans using both 2D and 3D scans. \textbf{2)Computer Aided Diagnosis(CADx)} that enhances malignancy classification by utilizing both imaging features and EMR data.
The main \textbf{contributions} of this work are as follows:
\begin{enumerate}
\item Development of a complete end-to-end pipeline combining CADe and CADx for automated lung nodule detection and its subsequent characterization.
\item Integration of a novel text prompt suite in SAM2 for lung nodule segmentation and detection in a zero-shot manner.
\item Generation of synthetic electronic medical records to provide a clinical context for more accurate diagnosis.
\item Development of a classification architecture that combines radiomic features with EMR data to improve classification performance.
\end{enumerate}


\section{Methodology}
\subsection{Dataset Curation and Pre-Processing} 
For the evaluation of our proposed method, we utilized the well-established Lung Image Database Consortium (LIDC) dataset~\cite{armato2011lung}, which comprises 1018 lung CT scans annotated by four expert radiologists. The radiologists have also provided their radiomic feature assessments, including subtlety, internal texture, spiculation, margin, lobulation, calcification, and sphericity, along with malignancy rating. LUNA16 ~\cite{setio2017validation}, is a subset of LIDC that consists of 888 scans and a majority consensus 1186 nodules. The inclusion criteria of LUNA16 for lung nodules is carried in our evaluation. While LIDC provides high-resolution CT and comprehensive imaging annotations, it lacks associated EMR data. To enhance clinical relevance, we generated synthetic EMRs based on the relationship between radiomic features and patient characteristics. Age distributions are tied to spiculation/malignancy scores. Malignancy probability increases with age, especially $>60$ years, for solid nodules~\cite{age}.  Gender probabilities are weighted by internal structure and lobulation. Lobulated nodules are more prevalent in males, while ground-glass opacities are more common in females~\cite{gender}. Smoking status is biased by spiculation and internal texture. Spiculated nodules show a strong association with smoking history ($p < 0.001$) in clinical cohorts~\cite{international2006survival}. Alcohol consumption rates are linked to malignancy/calcification patterns. Calcified nodules are typically benign, with lower smoking/alcohol associations~\cite{alcohol}. 

By combining radiological annotations and EMRs, our research bridges the gap between image analysis and patient-specific diagnostics. EMRs are scarce in publicly available datasets because of privacy issues, one of the main reasons for synthetic generation under certain constraints. The studies that have used EMRs along with their datasets are private, and their results can not be reproduced. While synthetic EMRs are designed with clinical relevance, real clinical records may slightly differ across different populations or over time. Furthermore, the EMRs generated are limited, while real-world clinical data contain a lot of other variables such as occupational exposures, family cancer history, prior cancer diagnosis, medication history, BMI, and metabolic factors, etc. Our approach is a foundation for multimodal studies combining imaging and clinical data for more personalized lung cancer assessment.

\subsection{Network Architecture} 
In this study, we developed a complete end-to-end pipeline for lung cancer screening that combines automated nodule detection and clinically-informed diagnosis of lung cancer in CT scans. Figure 1 illustrates the framework of our proposed architecture which consists of two stages. The nodule detection phase processes whole volumetric CT scans to identify potential nodule locations and generate precise segmentation of nodules. Once the nodules are detected and segmented, they are cropped from the complete CT scan and passed to the second phase for diagnosis. The nodule diagnosis phase combines nodule patches with radiomic feature descriptors and synthetically generated EMR data and gives malignancy classification as benign or malignant. The system is evaluated against state-of-the-art benchmarks using standard performance metrics, including sensitivity, specificity, and area under the receiver operator characteristics curve (AUC). An ablation study validates the architectural choices, and the system's performance is compared against existing state-of-the-art models. 

\begin{figure}[H]
\centering
\includegraphics[width=1\textwidth, height=7.5cm]{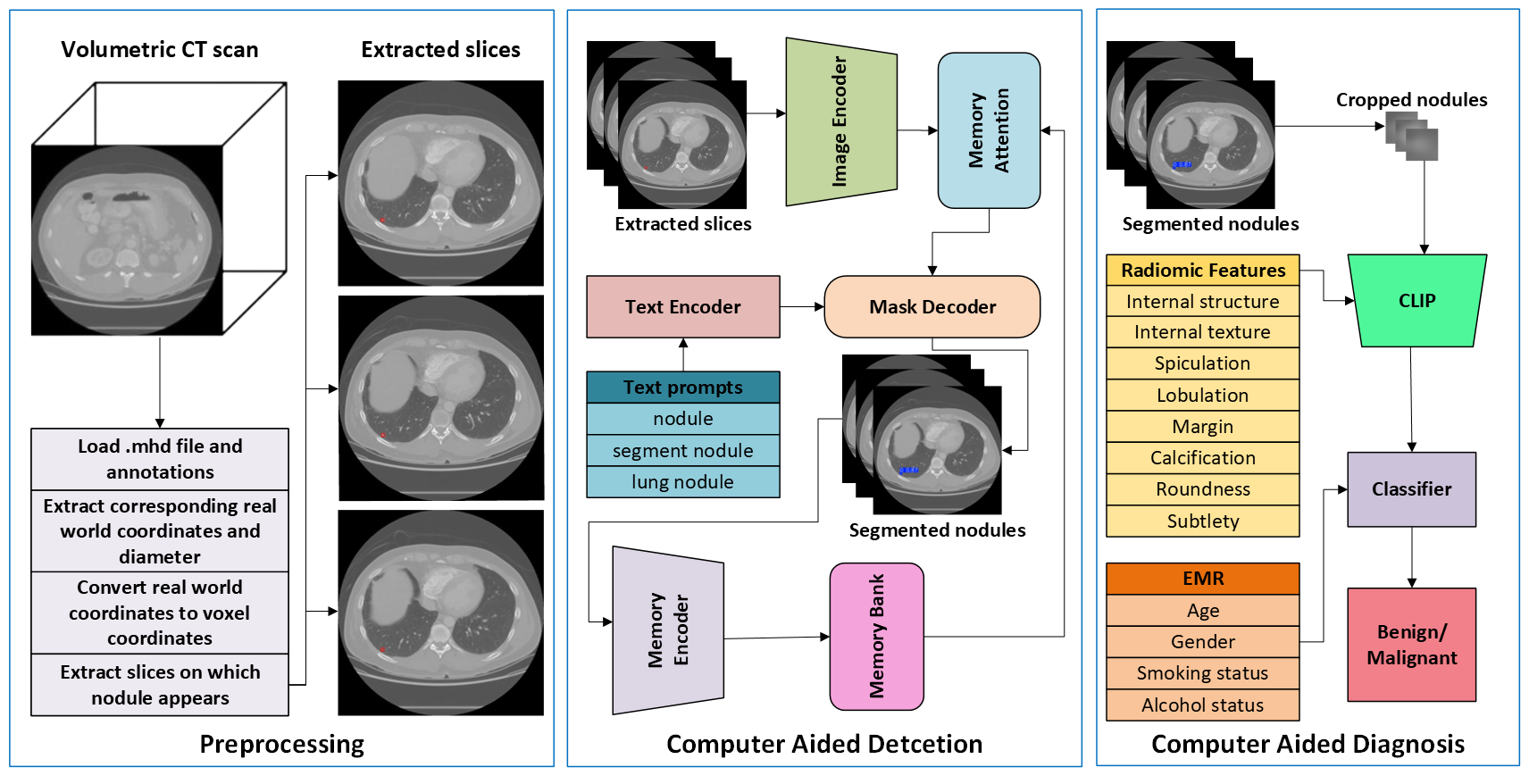}
\caption{Block diagram of the proposed framework. The initial stage is $CAD_{e}$, which detects nodule regions within the medical images. The subsequent stage, $CAD_{x}$, performs classification of the identified regions as benign/malignant.}\label{fig1}
\end{figure}

\paragraph{Nodule Detection (CADe):}  Recent advances in medical imaging have benefited significantly from the development of SAM2~\cite{ravi2024sam} and CLIP~\cite{clip} models. SAM2 achieves an improvement in processing speed while maintaining high segmentation accuracy and is capable of handling 3D data.  SAM2 is a prompt-based segmentation model that takes boxes, points, and masks as a prompt. These visual prompts require radiologists' intervention and increase processing time, making them less effective for clinical deployment. In our detection module, we have developed a novel text-prompt-based automated workflow. Our detection module introduces a novel, fully automated text-prompt-based detection model. Integrating a CLIP text encoder with SAM2 enables natural language prompts such as 'lung nodule' to segment nodules. The image encoder of the transformer generates high-dimensional image embeddings. The text encoder computes the similarity between the text prompt and the nodule region. The text prompt 'lung nodule' is converted to embeddings and is compared with image embeddings to compute the similarity score. The mask decoder combines image embeddings and text embeddings and generates high-resolution segmentation masks. The model was used in a zero-shot manner for lung nodule detection and segmentation. This fully automated workflow can overcome the time constraints in visual prompts and assist radiologists in efficiently segmenting nodules. With the proposed architecture, radiologists or clinicians can give the model natural language prompts such as 'lung nodule' to obtain the region of interest. The performance of our model was evaluated using standard segmentation metrics such as precision, recall, Dice Score, IoU, and compared against other state-of-the-art methods.

\paragraph{Nodule Classification (CADx):} In the second phase of our model, we have used CLIP~\cite{radford2021learning} for nodule classification. CLIP extracts high-dimensional image embeddings that capture the semantic characteristics of nodules. The similarity score between image embeddings and labels has been calculated using CLIP, which is then aligned with radiomic features, making an image-feature pair. These image-feature pairs, along with synthetic EMRs that include patient-specific data, are passed to the model for training. This combination of visual and clinical data helps improve diagnostic accuracy by providing both contextual and clinical information. ResNet50 ~\cite{he2016deep} is used as an image encoder. ResNet50 extracts spatial and texture features from the nodule images, capturing important patterns crucial for classification. The extracted features are fused with radiomic features, and cosine similarity is calculated to measure the closeness of the extracted features to predefined classes. The most similar class is given to the classifier along with EMR data for the final decision of nodules being benign or malignant. The model can make more accurate predictions by combining image embeddings, radiomic features, and clinical information. Finally, a classifier is trained on this comprehensive set of features to distinguish between benign and malignant nodules.

\begin{figure}[H]
\centering
\includegraphics[width=1\textwidth]{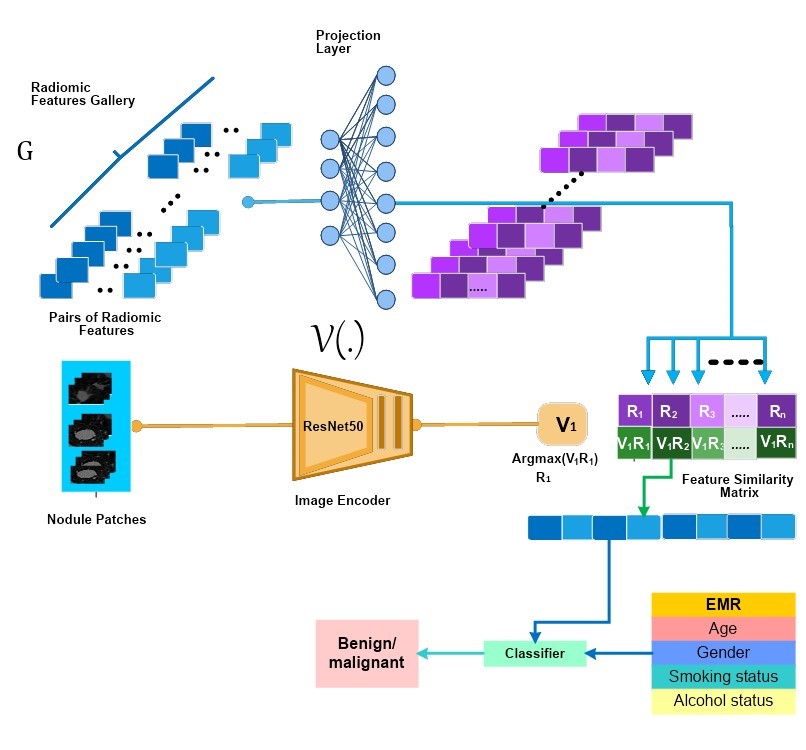}
\caption{Detailed architecture of the proposed method for lung nodule classification.} 
\label{fig2}
\end{figure}

We propose a comprehensive end-to-end pipeline that integrates both the detection and diagnosis of lung nodules. A key advantage of this approach is its ability to reduce or eliminate false positives identified during the detection stage through a subsequent diagnosis phase. We maintained high precision during detection to ensure potential nodules were not missed, while the diagnosis stage was designed to filter out false positives. The diagnosis phase used multimodal data for better classification results. Radiomic features and EMR provide additional contextual information for better classification accuracy. The pipeline integrates detection and diagnosis into a single system to improve efficiency and reduce the need for manual intervention. 

\subsection {Loss Function}
\paragraph{\textbf{Nodule Segmentation:}}
For the nodule segmentation task, we have used the same loss function as in MedSAM~\cite{ma2024segment}. For reference, the unweighted sum of cross-entropy loss and dice loss was selected because of their robustness and wide adaptation to medical image segmentation tasks. Specifically, let $S$ and $G$ represent the segmentation result and the ground truth, respectively. $s_i$, $g_i$ represent the predicted segmentation and ground truth of the voxel $i$, respectively. $N$ is the number of voxels in the image $I$, the binary cross-entropy loss is given as:

\begin{equation}
   L_{BCE} = -\frac{1}{N}\sum_{i=1}^{N}[g_ilogs_i+(1-g_i)log(1-s_i)],
\end{equation}

and dice loss is defined by

\begin{equation}
   L_{Dice} = 1-\frac{2\sum_{i=1}^Ng_is_i}{\sum_{i=1}^N(g_i)^2+\sum_{i=1}^N(s_i)^2},
\end{equation}

The final loss $L$ is defined by

\begin{equation}
   L = L_{BCE} + L_{Dice},  
\end{equation}
\vspace*{-\baselineskip}
\paragraph{\textbf{Nodule Classification:}}
For nodule classification as benign or malignant, we used the loss function defined in CLIP~\cite {radford2021learning} as:

\begin{equation}\label{lossfunc}
L_{\mathrm{SCE}} = \alpha \, L_{\mathrm{CE}}(p,q) + \beta \, L_{\mathrm{RCE}}(p,q),
\end{equation}

where $L_{\mathrm{CE}}(p,q) = -\sum_i p_i \log q_i$ denotes the cross-entropy loss, 
$L_{\mathrm{RCE}}(p,q) = -\sum_i q_i \log p_i$ denotes the reverse cross-entropy loss, 
$\alpha$ and $\beta$ are weighting coefficients for the two losses, 
and $p_i$ and $q_i$ denote the predicted and true probabilities, respectively.

\vspace*{-\baselineskip}
\section{Experimental Results and Discussions}
The experimental framework consists of two distinct stages. Initially, a detection model is developed to identify and segment areas containing nodules within the input CT scans, which can be either volumetric or slice-based. Subsequently, the segmented nodule regions are forwarded to a diagnosis model that determines the malignancy or benignity of the nodules. By unifying the detection and diagnosis processes, the system enhances both accuracy and reliability for improved early-stage detection and diagnosis.

For our nodule detection model, we employed zero-shot segmentation through the SAM2 model, complete with a ViT backbone, utilizing its standard architecture and pretrained weights. We transformed the medical images from MHD format to numpy arrays, normalizing the pixel values to a [0, 255] scale to ensure compatibility with the model. Dataset annotation coordinates in real-world space were translated to voxel coordinates using image origin and spacing metadata from the header files. To integrate text prompts, we used the CLIP text prompt encoder (ViT-B/32, 32x32 patches) in combination with SAM2, facilitating alignment of image and text features to calculate similarity scores for prompts like "lung nodules".

Evaluation of the segmentation model was conducted using both 2D- and 3D-forms of the LIDC-IDRI dataset. The model processed 3D volumetric CT scans, effectively accommodating the intricate spatial arrangement of lung nodules over numerous slices. For 2D analysis, the model's proficiency was assessed on individual slices to confirm its capability in slice-wise annotation. In both scenarios, segmentation relied on a bounding box along with a text prompt. Employing visual prompts, our segmentation model surpassed contemporary leading methods, achieving remarkable results. In 2D representation, it attained a Dice score of 0.92, an IoU of 0.85, a Precision of 0.91, and a Recall of 0.97, surpassing most current techniques in terms of segmentation accuracy and sensitivity. When using text prompts, there was a slight reduction in performance compared to visual prompts, but it still held competitive standing.

In 3D visualization, the model attained a Dice score of 0.81 and a recall of 0.77, indicating its robust ability to identify nodules using semantic guidance. Table 1 illustrates a comparison between our model and the current leading segmentation techniques using standard performance metrics. When compared to competitors such as MRUNet-3D, MDFN, and Wavelet U-Net++, our model exhibits superior segmentation performance. Remarkably, even in fully automatic text-prompted mode, the model surpasses certain conventional methods that depend on manual intervention and are either non-automatic or semi-automatic.

\begin{table}
\centering
\resizebox{\textwidth}{!}{ 
\begin{tabular}{l c c c c c c c c}
\toprule
\textbf{Author} & \textbf{Year} & \textbf{Method} & \textbf{Dataset} & \textbf{Samples} & \makecell{\textbf{Dice} \\ \textbf {Score}} & \textbf{IoU} & \textbf{Precision} & \textbf{Recall} \\
\midrule
\makecell{Babosa \\et al.\cite{bbosa2024mrunet}}  & 2024 & \makecell{MRUNet-\\3D} & LUNA16 & 888 & 0.83 & 0.86 & 0.83 & - \\
\makecell{Agnes et \\al.\cite{gao2024munet++}} & 2024 & \makecell{Wavelet\\ U-Net++} & \makecell{LIDC-\\IDRI} & 1018 & 0.93 & 0.87 & - & - \\
\makecell{Selvadass \\et al.\cite{selvadass2024satunet}}  & 2024 & \makecell{SAtUNet} & \makecell{LIDC-\\IDRI} & 1018 & 0.81 & 0.72 & 0.91 & 0.84 \\
\makecell{Cai \\et al.\cite{cai2024mdfn}}  & 2024 & \makecell{MDFN: A\\ Multi-level \\ Dynamic \\ Fusion\\ Network} & LUNA16 & 888 & 0.89 & 0.81 & - & 0.88 \\
\textbf{Ours} & \textbf{2025} & \makecell{\textbf{EMeRALDS} \\ (Visual \\ Prompt)} & \makecell{\textbf{LIDC-} \\ \textbf{IDRI(3D)}} & \textbf{1018} & \textbf{0.89} & \textbf{0.81} & \textbf{0.84} & \textbf{0.96} \\
\textbf{Ours} & \textbf{2025} & \makecell{\textbf{EMeRALDS} \\ (Text \\ Prompt)} & \makecell{\textbf{LIDC-}\\ \textbf{IDRI(3D)}} & \textbf{1018} & \textbf{0.81} & \textbf{0.70} & \textbf{0.89} & \textbf{0.77} \\
\textbf{Ours} & \textbf{2025} & \makecell{\textbf{EMeRALDS} \\ (Visual \\ Prompt)} & \makecell{\textbf{LIDC-} \\ \textbf{IDRI(2D)}} & \textbf{1018} & \textbf{0.92} & \textbf{0.85} & \textbf{0.91} & \textbf{0.97} \\
\textbf{Ours} & \textbf{2025} & \makecell{\textbf{EMeRALDS} \\ (Text \\ Prompt)} & \makecell{\textbf{LIDC-}\\ \textbf{IDRI(2D)}} & \textbf{1018} & \textbf{0.87} & \textbf{0.78} & \textbf{0.91} & \textbf{0.81} \\       
\bottomrule
\end{tabular}
} 
\caption{Comparison of our segmentation model's performance across different state-of-the-art methods on the LUNA16 and LIDC-IDRI datasets.}
\label{tab:segmentation_comparison}
\end{table}

Figure 3 illustrates several visual results generated by the CADe model on the LIDC-IDRI dataset for 2D imaging. It compares segmentation results obtained from visual prompts with those derived from text prompts. Each row presents a different CT scan, showing the ground truth along with the results predicted by the model. Furthermore, Figure 4 displays the 3D segmentation outcomes. Each row represents a distinct CT scan, encompassing the ROI, the ground truth, and the segmentation masks predicted by the model. These findings highlight the model's efficiency in identifying and segmenting nodules across various datasets.

\begin{figure}[t]
\centering
\includegraphics[width=1\textwidth]{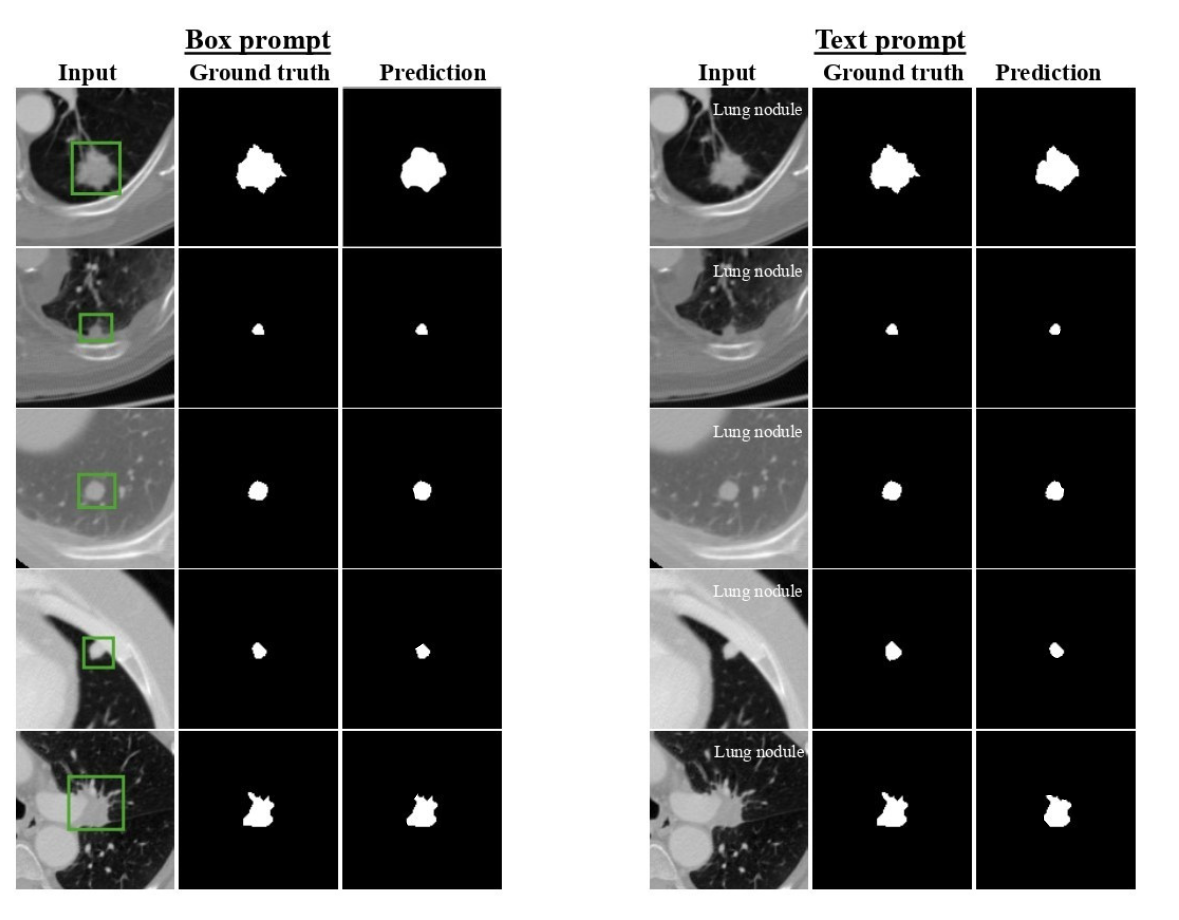}
\caption{Lung nodule segmentation results on LIDC-IDRI (2D slices) using box (left) and text (right) prompts. Each row shows the input CT slice, ground truth annotation, and predicted segmentation masks.} \label{LIDC}
\end{figure}

\begin{figure}[t]
\centering
\includegraphics[width=1\textwidth, height=9.5cm]{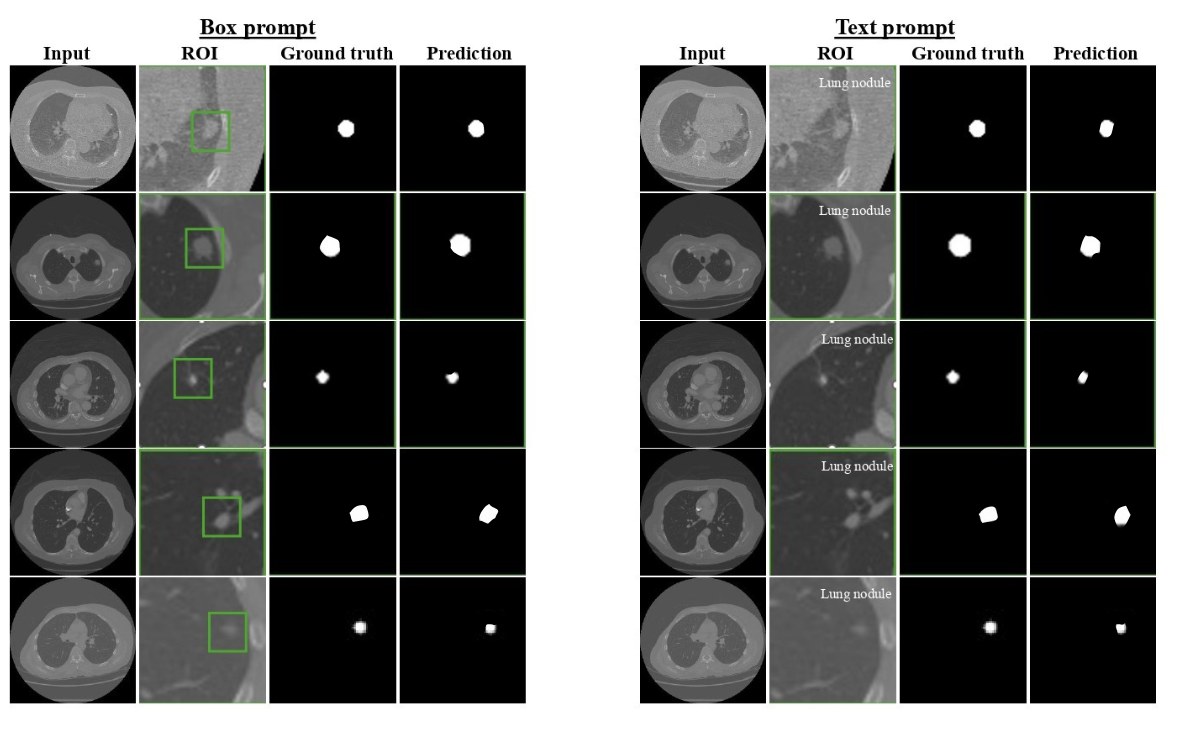}
\caption{Lung nodule segmentation results on LIDC-IDRI (3D CT scans) using box (left) and text (right) prompts. Each row displays an input CT slice, the corresponding ground truth annotation, and the predicted segmentation masks.}
\label{LUNA}
\end{figure}

In the nodule classification phase, the segmented nodule patches are input into the diagnosis model, which calculates a similarity score based on radiomic features. This score and synthetically generated clinical data (EMR) are then passed to a classifier for the binary classification of nodules as benign or malignant. Among the full set of radiomic features, only five are selected based on their importance to malignancy prediction. The similarity score and EMR features are fed into a classifier that performs binary classification as benign or malignant. Malignancy scores, ranging from 1 to 5, were converted into ground truth labels using a thresholding approach. Scores above 3 were labeled as malignant, scores below 3 were labeled as benign, and samples with an average score of exactly 3 were excluded from the evaluation to maintain a clear distinction between classes. The model was evaluated using standard metrics, including accuracy, precision, Recall, F1 score, specificity, and AUC-ROC, with a focus on minimizing false negatives to prioritize the early detection of malignant cases.

The proposed classification model, which combines radiomic features with electronic medical records (EMR), demonstrates superior performance compared to state-of-the-art methods. With an accuracy of 0.94 and an AUC of 0.95, the model outperforms existing approaches such as TransUNET (0.85 accuracy, 0.86 AUC), Swin-T variants (0.72–0.82 accuracy), and Vision Transformers (ViT) (0.79 accuracy, 0.79 AUC). Furthermore, the model achieves a high specificity of 0.97, indicating its strong ability to correctly identify negative cases, which is crucial in clinical diagnostics. Recall (0.72) suggests that there is still room for improvement in detecting all true positive cases. However, the integration of EMR data significantly improves diagnostic performance, as shown in Table 2 by the model's performance. These results highlight the advantage of combining radiomic features with electronic medical records (EMRs), underscoring the potential of multimodal approaches in medical diagnosis.

\begin{table}[t]
    \centering
    \resizebox{\textwidth}{!}{ 
    \begin{tabular}{l c p{3cm} p{2cm} c c c c c}
        \toprule
        \textbf{Classifier} & \textbf{Accuracy} &  \textbf{Precision} & \textbf{Recall} & \textbf{F1} & \textbf{AUC} & \textbf{Specificity} \\
        \midrule
        TransUNET~\cite{article2} & 0.85 & - & 0.71 & - & 0.86 & 0.93\\
        Swin-T + Swin-B ~\cite{sun2022efficientlungcancerimage} & 0.82 & - & - & - & - & -\\
        SwinT~\cite{Fan2024TStage} & 0.72 & 0.79 & 0.80 & 0.73 & 0.93 & -\\
        HCT~\cite{tomography10100123} & 0.83 & - & 0.72 & - & 0.73 & 0.84 & \\
        ViT~\cite{tomography10100123} & 0.79 & - & 0.71 & - & 0.79 & 0.79\\
\textbf{Ours (Radiomic Features + EMR)} & \textbf{0.94} & \textbf{0.84} & \textbf{0.72} & \textbf{0.77} & \textbf{0.95} & \textbf{0.97} \\
\textbf{Ours (Radiomic features only)} & \textbf{0.86} & \textbf{0.88} & \textbf{0.67} & \textbf{0.76} & \textbf{0.91} & \textbf{0.95} \\
        \bottomrule
    \end{tabular}
    } 
    \caption{Comparison of our classification model's performance with and without EMR across different state-of-the-art methods.}    
    \label{tab:segmentation_comparison2}
\end{table}

Different classifiers were trained using the selected radiomic features and EMR data. Among these models, Gradient Boosting achieved the highest overall performance, with an accuracy of 0.94, F1-score of 0.77, and AUC of 0.95. The Cubic SVM, Medium KNN, and Linear SVM also performed competitively, though with slightly lower F1-scores and precision. The superior performance of Gradient Boosting suggests it is well-suited for handling heterogeneous feature sets in medical datasets. This is supported by the AUC curves in Figure 5, which compare the AUC of the top four models. Gradient Boosting achieved the highest AUC (0.95), followed by Random Forest (0.94), and SVM with cubic and linear kernels (both 0.93). The curve visually demonstrates the model's ability to distinguish between classes. The AUC values confirm that all models perform well, but Gradient Boosting is the most reliable for this classification task.

\begin{figure}[H]
\centering
\includegraphics[width=0.75\textwidth]{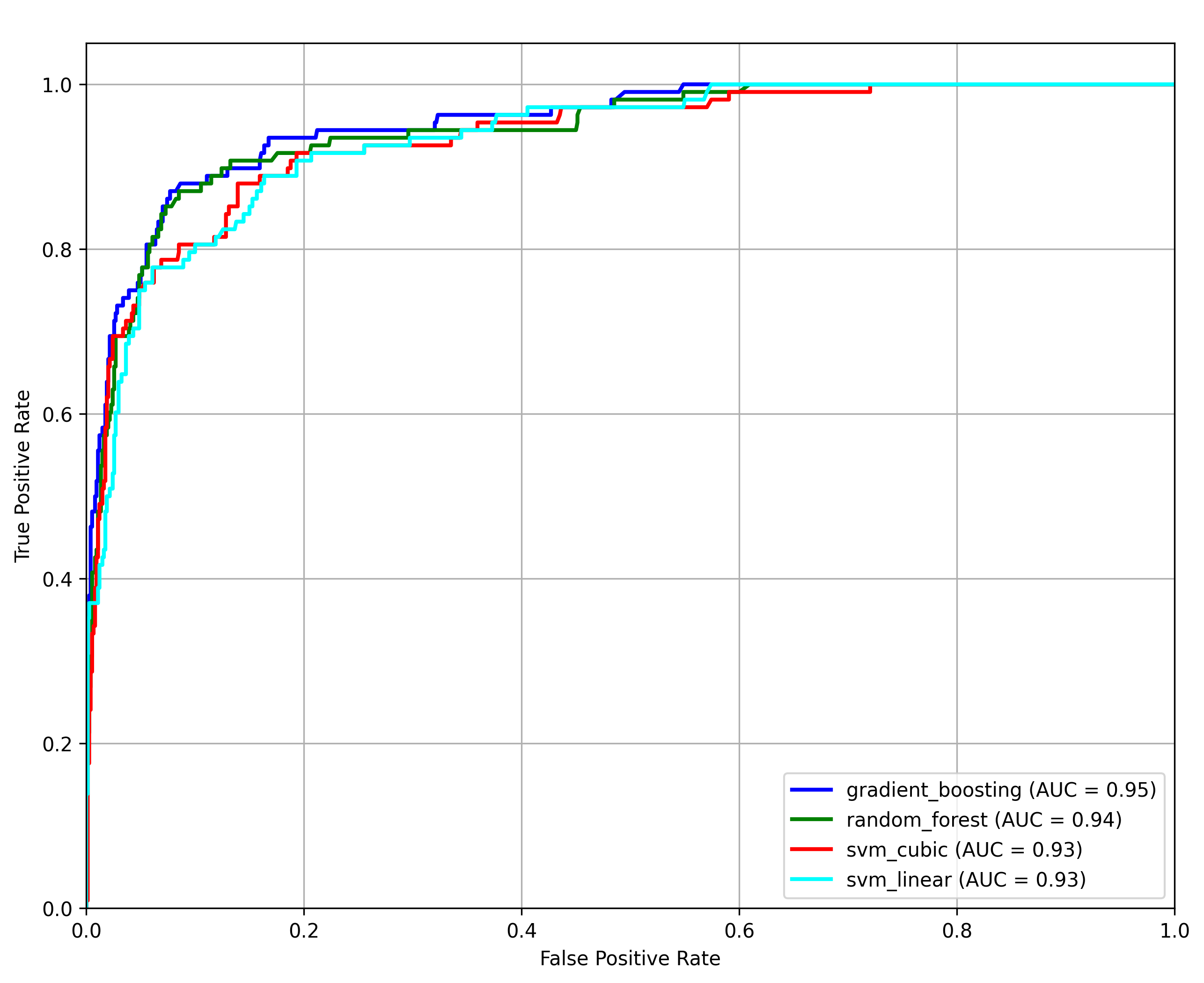}
\caption{AUC curve of top four models.} \label{LUNA2}
\end{figure}

\noindent{\textbf{Ablation study:}} The ablation study investigates the influence of varying the number of radiomic features on the model's performance across several key evaluation metrics. Figure 6 presents the results for precision, F1-score, AUC, and accuracy as functions of the number of radiomic features used in the model. The study reveals that using five radiomic features consistently provides the best performance when compared to other feature subsets.

Figure 6(a) shows the relationship between precision and the number of features, indicating that the highest value is achieved when five features are used. Similarly, Figure 6(b) demonstrates the F1-score with k number of features, where the highest F1-score is observed with five features. In Figure 6(c), the AUC (Area Under the Curve) shows the peak performance at five features. The AUC provides insight into the model's ability to distinguish between benign and malignant nodules, and the results suggest that five features provide an optimal balance of sensitivity and specificity, leading to better overall model performance. Lastly, Figure 6(d) highlights accuracy, which also reaches its peak when five features are used, reinforcing the choice of five as the most effective feature subset for this classification task.

\begin{figure}[t]
\centering
\includegraphics[width=1\textwidth, height=11cm]{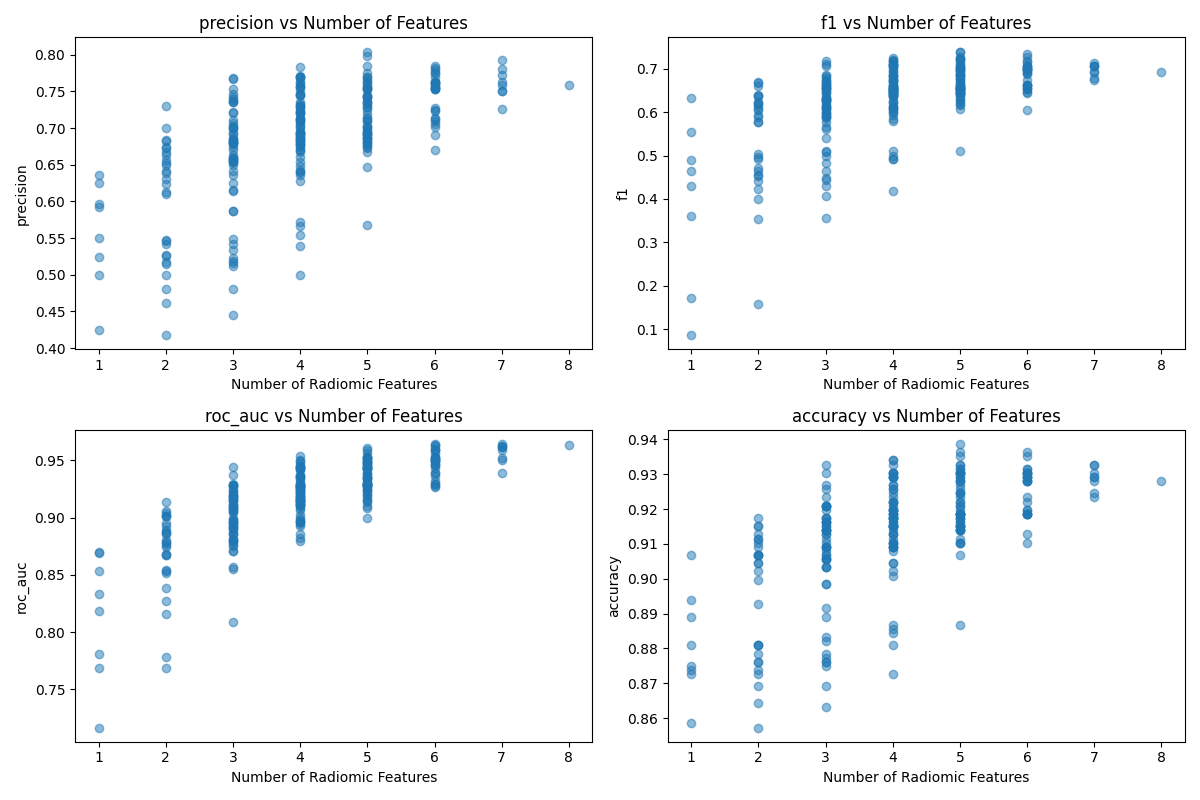}
\caption{Different evaluation parameters on k number of radiomic features (a) precision vs number of features (b) F1 score vs number of features (c) ROC vs number of features (d) Accuracy vs number of features.} \label{LIDC2}
\end{figure}

Furthermore, to evaluate the individual contribution of clinical data, an ablation study was conducted using only the radiomic features, excluding Electronic Medical Records (EMR), as input to the classification models. Figure 7(a) illustrates the performance of the different classifiers when trained on the five selected radiomic features alongside EMRs, while Figure 7(b) demonstrates the performance of the same classifiers trained on the selected five radiomic features without EMRs.

For instance, the Gradient Boosting model, which previously achieved an accuracy of 0.94 with combined features, showed a reduced accuracy of 0.86. Other classifiers, such as Cubic SVM and Linear SVM, also exhibited noticeable declines in precision, recall, and AUC, indicating that clinical variables provide complementary information that enhances the classification ability of the models. Clinical attributes, such as patient age, smoking history, and gender, can influence the malignancy risk and offer context that radiomic features alone may not fully capture. This underscores the importance of clinical data fusion in medical AI applications.

A comparison shown in Figure 7 reveals a consistent performance drop when EMR data is omitted. Without EMR, the model relies solely on radiomic features, resulting in a reduction in accuracy from 0.94 to 0.86 and a drop in AUC from 0.95 to 0.91, indicating diminished discriminative power. The recall falls from 0.72 to 0.67, suggesting that the absence of EMR leads to the missed detection of positive cases, a critical drawback in medical diagnosis. Additionally, specificity decreases slightly from 0.97 to 0.95, meaning the model becomes marginally less effective at correctly ruling out negative cases.  

This performance degradation highlights that radiomic features alone, while informative, lack the clinical insights provided by EMR. The decline in recall is particularly concerning, as it implies higher false-negative rates, which is a significant risk in clinical settings where early detection is crucial. The results suggest that EMR data helps bridge gaps in image-based analysis, improving both sensitivity and overall diagnostic reliability. The findings reinforce that multimodal approaches, combining imaging and clinical data, are essential for robust medical AI systems.

\begin{figure}[H]
\centering
\includegraphics[width=1\textwidth]{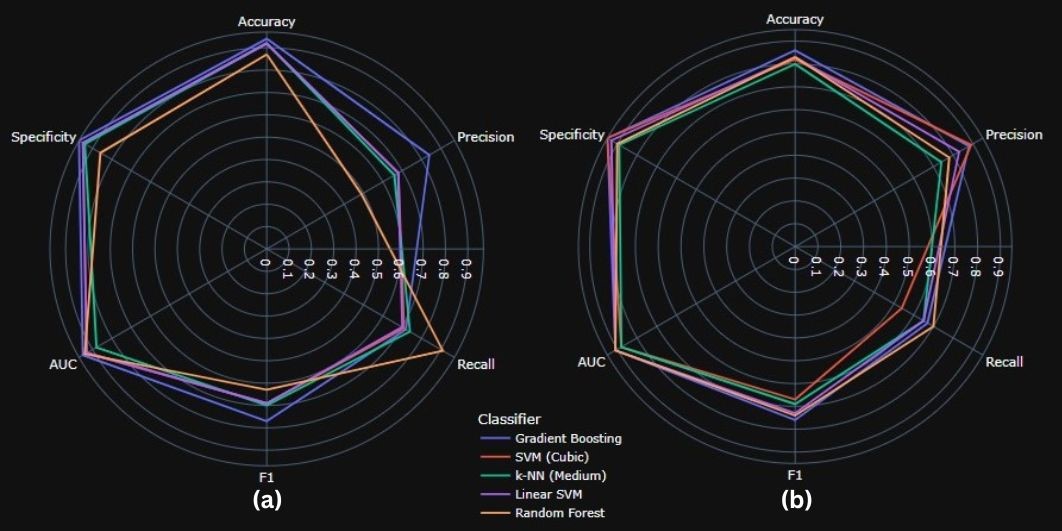}
\caption{Radar chart showing the performance comparison of classifiers for lung nodule malignancy prediction using (a) selected radiomic features and EMR (b) selected radiomic features only (without EMRs).} \label{LIDC3}
\end{figure}

\noindent{\textbf{Model limitations:}} A primary limitation of this study is the reliance on a limited annotated dataset with weak labeling. While weak labels can still provide useful information, their inherent noise can limit the accuracy and robustness of the model. Another limitation is the use of synthetic EMRs for the classification phase. Although synthetic EMRs are useful for simulating clinical scenarios, they fall short of capturing the full complexity and variability of real-world data. Incorporating real-time clinical data could enhance the system's performance. As a future direction, integrating real-world EMRs into the diagnostic pipeline may further unlock the potential of these multimodal models in practical clinical applications. Integrating real-time, real-world EMRs would provide more accurate demographic, clinical, and diagnostic information, which could significantly improve the model's predictive capabilities. 

\section{Conclusion}
In this study, we introduced an end-to-end pipeline that integrates computer-aided detection and computer-aided diagnosis for analyzing lung nodules, aiming to assist in early lung cancer screening. The pipeline utilizes SAM2 for zero-shot lung nodule segmentation, enhanced with text prompts using the CLIP text encoder to guide the segmentation process. For the classification phase, we utilized a modified CLIP-based architecture trained on segmented nodule patches and their corresponding radiomic feature sets, enabling robust representation learning. During inference, the model determined the most similar radiomic class based on feature similarity. This information, along with synthetic clinical data (EMRs), was then fed into a classifier for the final benign/malignant classification. Our experimental results demonstrate that integrating radiomic features and clinical context significantly improves diagnostic performance, and the use of foundation models allows for effective generalization across diverse imaging conditions. The proposed pipeline presents a promising direction for scalable, automated lung cancer screening, with potential for real-world deployment. Future improvements can focus on enhancing label quality and incorporating actual EMRs to further validate clinical applicability.

\section{Acknowledgements}
\label{app1}
This work is part of the NRPU project $\#$ 17019 entitled “EMeRALDS: 
Electronic Medical Records driven Automated Lung nodule Detection and cancer risk Stratification” funded by Higher Education Commission of Pakistan.
\section{Ethics Statement}

This study did not involve any experiments on humans or animals conducted by the authors. 
The research was conducted using publicly available, anonymized chest CT datasets, including 
the Lung Image Database Consortium (LIDC-IDRI) and the LUNA16 dataset. These datasets 
contain no identifiable personal information, and all data collection and sharing procedures 
complied with relevant ethical and legal guidelines. Therefore, separate institutional ethics 
approval was not required for this study.

\bibliographystyle{elsarticle-harv} 
 \bibliography{manuscript}
\end{document}